\documentclass[twocolumn,aps,prl,epsfig,graphics]{revtex4}%
\usepackage{graphicx}
\usepackage{amsmath}
\usepackage{amsfonts}
\usepackage{amssymb}%
\begin{document}
\title{Interfacing quantum optical and solid state qubits}
\author{L. Tian$^{1}$, P. Rabl$^{1}$, R. Blatt$^{2}$ and P. Zoller$^{1}$}

\affiliation{$^{1}$ Institute for Theoretical Physics, University of Innsbruck,
A--6020 Innsbruck, Austria\\
$^{2}$ Institute for Experimental Physics, University of
Innsbruck, A--6020 Innsbruck, Austria }

\begin{abstract}
We present a generic model of coupling quantum optical and solid
state qubits, and the corresponding transfer protocols. The example
discussed is a trapped ion coupled to a charge qubit (e.g. Cooper
pair box). To enhance the coupling, and achieve compatibility
between the different experimental setups we introduce a
superconducting cavity as the connecting element.
\end{abstract}
\maketitle

%\address{$^{1}$ Institute for Theoretical Physics, University of Innsbruck,
%A--6020 Innsbruck, Austria}

Significant progress has been made during the last few years in implementing
quantum computing proposals with various physical
systems\cite{divincenzo_exp_quant_comp_2000}. Prominent examples are quantum
optical systems, in particular trapped ions\cite{cz_gate} and atoms in optical
lattices\cite{Jaksch_rydberg}, and more recently solid state systems, such as
Josephson junctions\cite{charge_qubit_schoen,flux_qubit_mooij} and quantum
dots\cite{loss_divincenzo_dots}. Among the recent experimental highlights are
demonstration of, or first steps towards coherent control and read out of
qubits, engineering of entanglement, and simple quantum
algorithms\cite{cnot_two_ions_innsbruck,deutsch_jozsa_innsbruck_2003,cnot_one_ion_nist,nakamura_two_bits,delft_two_bit_gates}%
. Based on this progress, the question arises to what extent
hybrid systems\cite{cpb_mechanical_resonator_schwab}
can be developed with the goal of combining advantages of various approaches,
while still being experimentally compatible. At the heart of this problem are
questions of interfacing quantum-optical qubits and solid state systems: this
can be either in a form where quantum-optical qubits are connected via a solid
state data bus as a route to scalable quantum computing, or by reversibly
transferring a quantum optical to solid state qubits and circuits.

Below we will discuss a generic physical model and corresponding protocol for
interfacing a quantum-optical and solid state charge qubit. On the quantum
optics side, qubits are typically represented by long lived internal (nuclear)
spin states of single atoms or ions\cite{ion_trap_decoherence_review}. These
single atoms can be trapped and laser cooled. External fields,
provide a mechanism to manipulate the qubit, as well as to entangle the
qubit with the motional state of the trapped particle. In the
case of trapped ions this allows one to convert spin (the qubit) to charge
superpositions, either dynamically, e.g. by kicking with a laser, or
quasi-statically by applying spin-dependent (optical or magnetic) potentials.
This spin to charge conversion provides a natural capacitive coupling to a
solid state charge qubit, represented e.g. by a Josephson
junction\cite{charge_qubit_schoen} or a charged double quantum
dot\cite{double_dot_rmp}. Instead of direct coupling of the charges, one can
introduce auxiliary elements, such as cavities. This serves the purpose of
allowing increased spatial separation and mutual shielding of the systems,
with the goal of easing experimental requirements for coexistence of the
hybrid qubits (e.g. trapping and laser manipulation of atoms or ions), while
enhancing the coupling strength in comparison with free space. Features of
this hybrid system are the significantly different time scales of the evolution
and coupling of the two qubits, and (in comparison with quantum optical
qubits) short decoherence time of the solid state systems. We exploit this by
developing a protocol of a ``fast swap'' with kicking by short pulses on
a time scale comparable to the charge - charge coupling which is much shorter
than the trap period. This has the additional benefit of being a hot gate,
i.e. not requiring cooling to the motional ground state, and not making a
Lamb-Dicke assumption of strong confinement.

\textit{Model for coupled qubits.} A Hamiltonian for our combined system has
the form $H_{t}=H_{s}+H_{q}+H_{\mathrm{int}}$ with Hamiltonian for the quantum
optical qubit
\begin{equation}
H_{s}=\left(  \frac{\hat{p}_{x}^{2}}{2m}+\frac{1}{2}m\omega_{\nu}^{2}\hat
{x}^{2}\right)  +\frac{\delta_{0}}{2}\sigma_{z}^{s}+\left(  \frac{\hbar
\omega_{R}(t)}{2}\sigma_{+}^{s}e^{i\delta k_{l}\hat{x}}+\mathrm{h.c.}\right)
. \label{Hs}%
\end{equation}
the solid state charge qubit,
\begin{equation}
H_{q}=\frac{E_{z}}{2}\sigma_{z}^{q}+\frac{E_{x}}{2}\sigma_{x}^{q} \label{Hq}%
\end{equation}
and interaction term
\begin{equation}
H_{\mathrm{int}}=\hbar\kappa(t)\hat{x}\sigma_{z}^{q}. \label{Hint}%
\end{equation}

The first term in $H_{s}$ describes the $1$D-motion of a charged particle
(ion) in the harmonic trapping potential with $\hat{x}$ the coordinate,
$\hat{p}_{x}$ the momentum and $\omega_{\nu}$ the trapping frequency.
A pseudo-spin notation with Pauli operators $\sigma_{i}^{s}$ describes
the atomic qubit. Physically, the qubit is represented by two atomic ground
state levels which are coupled by a laser induced Raman transition with Rabi
frequency $\omega_{R}(t)$ and detuning $\delta_{0}$. Transitions between the
states are associated with a momentum kick $\delta k_{l}$ due to photon
absorption and emission, which couples the qubit to the motion at
the Rabi frequency $\omega_{R}(t)$\cite{juanjo_fast_gate}.

\begin{figure}[tbh]
\includegraphics[width=6cm]{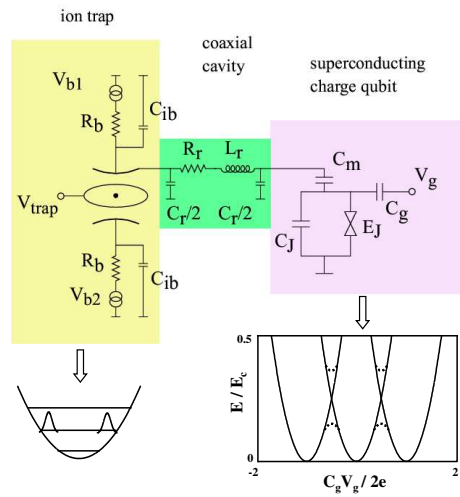}\caption{Schematic
coupling circuit of the charge qubit and trapped ion. Top: the
coupling via a cavity. The voltage at the electrode is balanced by
a filtering circuit. Bottom left: coherent states of the motional
mode. Bottom right:
energy of charge qubit vs gate voltage.}%
\label{figure_1}%
\end{figure}

The Hamiltonian for the solid state charge qubit has the generic form
(\ref{Hq}) with $\sigma_{i}^{q}$ the Pauli operators, and $E_{z}$ and $E_{x}$
being tunable parameters. A Hamiltonian of this form is obtained, for example,
for a superconducting charge qubit, i.e. a
superconducting island connecting with a low capacitance and high resistance
tunnel junction (see Fig.~\ref{figure_1}(a)). With a phase
$\varphi$ and its conjugate $\hat{n}$, the Hamiltonian is $H_{q}=E_{c}(\hat
{n}+C_{g}V_{g}/2e)^{2}-E_{J}\cos{\varphi}$\cite{charge_qubit_schoen}, where
$E_{J}$ is the Josephson energy and $E_{c}$ is the capacitive energy with
$E_{c}\gg E_{J}$. The gate voltage $V_{g}$ controls the qubit through the gate
capacitor $C_{g}$. The qubit forms an effective two level system with charge states
$|0\rangle=|n\rangle$ and $|1\rangle=|n+1\rangle$, and
$E_{z}=E_{c}(C_{g}V_{g}/2e)$ and $E_{x}=E_{J}$. Adjusting $V_{g}$ or $E_{J}$
provides arbitrary single qubit gates. Typically, $E_{c}$ is about
$100\,\mathrm{GHz}$ and $E_{J}$ is about $10\,\mathrm{GHz}$. Other solid-state
systems such as a double quantum dot qubit can be considered within a
similar framework.

Finally, the interaction (\ref{Hint}) has the universal form of a linear
coupling via the coordinate $\hat{x}$ and the charge operator
$\sigma_{z}^{q}$ which characterizes the electrostatic coupling between the
motional dipole and the charge. Note that this coupling is of the order
of $ep_{i}/4\pi\epsilon_{0}r_{0}^{2}$ for a given distance $r_{0}$ with dipole
$p_{i}$ and charge $e$, and is a factor of $er_{0}/p_{i} $ stronger
than the familiar dipole-dipole couplings encountered in quantum optics.
Instead of a direct coupling of the dipole to the charge, we introduce an
interaction via a short superconducting cavity. This provides a mutual
shielding of the qubits, e.g. from stray photon exciting quasiparticles which
might impair the coherence of the charge qubit. Furthermore, this allows the
interaction to be controlled by inserting a switch, e.g. a controllable
Josephson junction (a SQUID).

\textit{Coupling via a superconducting cavity.} The transmission line is
described by a phase variable $\psi(z,t)$ \cite{transmission_line_ref}, where
$z$ is the longitudinal direction of the cavity. The Lagrangian is
\begin{equation}
\mathcal{L}=\frac{C_{r}}{2L}\int_{0}^{L}dz\dot{\psi}^{2}-\frac{L}{2L_{r}}%
\int_{0}^{L}dz(\frac{\partial\psi}{\partial z})^{2}%
,\label{lagrangian_transmission_line}%
\end{equation}
with $C_{r}$ the capacitance of the cavity, $L_{r}$ the inductance,
and $L$ is the length. The modes for a finite length
transmission line cavity are the Fourier components of $\psi(z,t)=\sqrt{2}\sum
\psi_{n}(t)\cos(kz)$ with $k=(2n+1)\pi/L$ for open boundary condition and $n$
integer. When the cavity is made of two parallel cylindrical rods with a
distance $d_{0}$ and the rod radii $b_{0}$, the inductance is $L_{r}=\mu
_{0}\ln{(d_{0}/b_{0})}L/\pi^{3}$ and the capacitance is $C_{r}=4\pi
\epsilon_{0}L/4\ln{(d_{0}/b_{0})}$. For example, for $d_{0}=20\,\mathrm{\mu
m}$, $b_{0}=1\,\mathrm{\mu m}$, $L=100\,\mathrm{\mu m}$, which gives
$C_{r}=1\,\mathrm{fF}$, $L_{r}=10\,\mathrm{pH}$, the frequency of the first
eigenmode of the cavity $\omega_{r}/2\pi=1.5\,\mathrm{THz}$. Application of
millimeter transmission lines in the microwave regime has been proposed for
the interaction of charge qubits where the cavity mode is in resonance with
the qubit\cite{schoelkopf_coupling}.

The coupling scheme is shown in Fig.~\ref{figure_1}(a). The left end of the
cavity capacitively couples with the ion as $(V_{i}-\dot{\psi}(0,t))e\hat
{x}/d_{i}$, where $d_{i}$ is the distance between the ion and the cavity
and $V_{i}$ is the voltage on trap electrode. The cavity couples with one of the trap
electrodes by the capacitor $C_{i}$. The right end of the cavity couples with
the charge qubit via a contact capacitor $C_{m}$ as $C_{m}%
(\dot{\psi}(L,t)-\dot{\varphi})^{2}/2$. In our scheme, the cavity length is
much shorter than the wave length of a microwave field that is
characteristic of the energy of the charge qubit, so that the cavity can be
represented by phases $\psi_{1,2}$ at the ends of the cavity.
Each node is connected with the ground by a capacitor $C_{r}/2$, and the two
nodes are connected by the inductor $L_{r}$, as is shown in
Fig.~\ref{figure_1}(a). The conjugates of the phases obey the charge
conservation relation $p_{1}+p_{2}=0$, where the momentum operators
$p_{1,2}$ are the total charge on the nodes. With the new
variable $\tilde{\psi}=\psi_{1}-\psi_{2}$ and its conjugate $\tilde{p}_{\psi
}=(p_{1}-p_{2})/2$, the interaction is $H_{\mathrm{int}}=H_{\mathrm{cav}%
}+H_{1}$ with
\begin{equation}%
\begin{array}
[c]{c}%
\displaystyle H_{\mathrm{cav}}=\frac{\tilde{p}_{\psi}^{2}}{2(C_{r}/4)}%
+\frac{\tilde{\psi}^{2}}{2L_{r}}\\[4mm]%
\displaystyle H_{1}=\tilde{p}_{\psi}\frac{e\hat{x}/d_{i}+C_{i}V_{i}}%
{C_{i}+C_{r}/2}-\tilde{p}_{\psi}\frac{C_{m}}{C_{t}}\frac{p_{\varphi}%
+C_{g}V_{g}}{C_{m}+C_{r}/2}%
\end{array}
\label{motion_charge_Hamiltonian}%
\end{equation}
where $C_{t}=C_{m}+C_{J}+C_{g},$ and we have assumed $C_{r}\gg C_{i}%
,C_{m},C_{J},C_{g}$. The equation includes the Hamiltonian of the cavity
and the coupling between the cavity, the charge and the motion. The cavity
mode is an oscillator with the eigenfrequency $\omega_{r}=2/\sqrt{L_{r}C_{r}}%
$. With a second order perturbation approach and replacing $p_{\varphi}$ with
$e\sigma_{z}^{q}$, we derive the effective coupling between the charge qubit
and the ion motion,
\begin{equation}
H_{int}^{(2)}=\frac{e^{2}}{C_{r}}\frac{C_{m}}{C_{t}}(\frac{\hat{x}}{d_{i}%
}+\frac{C_{i}V_{i}}{e})(\sigma_{z}^{q}+\frac{C_{g}V_{g}}{e}),
\label{int_i_t_c}%
\end{equation}
where the coupling includes the effect of the gate voltage $V_{g}$
on the ion which shifts the trapping potential, and the effect of
the trap voltage $V_{i}$ on the charge qubit which can be avoided
by designing a balance circuit (below). The cavity shortens the
distance between the charge and the ion to the order of $d_{i}$
with $\hbar\kappa=e^{2}/2C_{r}d_{i}$ where $\kappa$ is increased
by a factor of $4\ln (d_{0}/b_{0})r_{0}/d_{i}$ compared with the
direct coupling. With $C_{g}\sim C_{J}=0.1\,\mathrm{fF}$ and
$C_{i}\sim C_{m}=0.2\,\mathrm{fF}$,
$\kappa/2\pi=25\,\mathrm{GHz}/d_{i}$.

\textit{A ``fast swap'' gate.} A controlled phase gate, $|\epsilon_{1}%
\rangle_{s} |\epsilon_{1}\rangle_{q} \rightarrow(-1)^{\epsilon_{1}
\epsilon_{2}} |\epsilon_{1}\rangle_{s}|\epsilon_{1}\rangle_{q}$ ($\epsilon
_{1,2}=0,1$), together with single qubit rotations forms a universal set of
operations required for entanglement and information exchange between the
ion and the charge. The swap gate, which is the key step for interfacing
the ion and the charge qubit, can be achieved by three
controlled phase gate together with Hadamard gates on the
qubits\cite{Nielsen_Chuang_book}.

We construct a phase gate operating on nanoseconds, a time scale much shorter than the trap
period, and not requiring the cooling of the phonon
state with a eight-pulse sequence. Three evolution operators are used in this
sequence. The free evolution $U_{0}(t)=\exp{(-i\omega_{\nu}t\hat{a}%
_{x}^{\dagger}\hat{a}_{x})}$, where $\hat{a}_{x}$ ($\hat{a}_{x}^{\dagger}$)
are the creation (annihilation) operator of the motion, is achieved by turning
off the interactions $\kappa=\omega_{R}=0$ and the energy $H_{q}$.
Entanglement between the motion and the spin state is obtained by applying
short laser pulses with $\omega_{R}\tau_{s}=\pi$ for $n_{l}$ times
($\kappa=H_{q}=0$ and $n_{l}$ even): $U_{l}(z_{l}n_{l})=\exp{(-iz_{l}\delta
k_{l}n_{l}\sigma_{z}^{s}\hat{x})}$, where $z_{k}=\pm1$ is the direction of the
photon wave vector and $\delta k_{l}$ is the momentum from one kick of the
laser. The duration of the kicking $\tau_{s}$ is assumed to be much shorter
than other time scales. Entanglement between the motion and the charge qubit
is obtained by turning on the interaction $\kappa$ for time $\tau_{q}$
($H_{q}=\omega_{R}=0$ and $\omega_{\nu}\tau_{q}\ll 1$): $U_{2}(\tau_{q}%
)=\exp{(-i\kappa\tau_{q}\sigma_{z}^{q}\hat{x})}$. By flipping the charge
qubit with single-qubit operation, the sign of the evolution can be flipped as
$U_{2}(-\tau_{q})=\sigma_{x}^{q}U_{2}(\tau_{q})\sigma_{x}^{q}$.

The gate sequence is
\begin{equation}%
\begin{array}
[c]{c}%
\displaystyle U(T)=U_{l}(n_{l}^{2})U_{q}(\tau_{q}^{2})U_{0}(t_{2})U_{l}%
(-n_{l}^{1}-n_{l}^{2})U_{q}(-\tau_{q}^{1}-\tau_{q}^{2})\\[2mm]%
\times U_{0}(t_{1})U_{l}(n_{l}^{1})U_{q}(\tau_{q}^{1})
\end{array}
\label{eight_pulse_gate_1}%
\end{equation}
where the parameters fulfill $n_{l}^{1}t_{1}=n_{l}^{2}t_{2}$ and $\tau
_{q}^{1}t_{1}=\tau_{q}^{2}t_{2}$. For a free particle with mass $m$ where
$U_{0}(t)=\exp{(-i\hat{p}^{2}t/2m\hbar)}$, the evolution is equivalent to%
\begin{equation}%
\begin{array}
[c]{c}%
\displaystyle U(T)=e^{-i(\hat{p}+\hbar k_{\mathrm{eff}}^{2})^{2}t_{2}/2m\hbar
}e^{-i(\hat{p}-\hbar k_{\mathrm{eff}}^{1})^{2}t_{1}/2m\hbar}\\[2mm]%
\displaystyle=e^{i\phi^{\prime}}U_{0}(t_{1}+t_{2})e^{-i\alpha\sigma_{z}%
^{q}\sigma_{z}^{s}}%
\end{array}
\label{free_particle}%
\end{equation}
where $\phi^{\prime}$ is a global phase, $k_{\mathrm{eff}}^{i}=\delta
k_{l}n_{l}^{i}\sigma_{z}^{s}+\kappa\tau_{q}^{i}\sigma_{z}^{q}$ with
$i=1,2$, and $\alpha=\hbar\kappa\delta k_{l}\tau_{q}^{1}n_{l}^{1}t_{1}%
/mt_{2}(t_{1}+t_{2})$. The motional part factors out from the evolution
of the qubits. Hence, the gate does not depend on the initial state of the
phonon mode. For an oscillator by
making the approximation $\exp{(-i\omega_{\nu}t)}\rightarrow1-i\omega_{\nu}t$,
the same result is obtained with the motional part replaced by $U_{0}%
(t_{1}+t_{2})\rightarrow\exp{(-i\omega_{\nu}(t_{1}+t_{2})\hat{a}_{x}^{\dagger
}\hat{a}_{x})}$. The fidelity is $1-O(\omega_{\nu}^{2}t^{2})$ (the gate is
exact for free particle) and can be increased by exploiting a low trapping
frequency. For the phase gate $\alpha=\pi/4$. This gives the total gate time%
\begin{equation}
T=\frac{\pi m}{4\hbar\kappa\delta k_{l}}(\frac{1}{n_{l}^{1}t_{1}}%
+\frac{1}{n_{l}^{2}t_{2}})+t_{1}+t_{2}\label{gate_time}%
\end{equation}
which shows that the limits of the phase gate are essentially set
by the available Rabi frequency of the laser and the coupling
$\kappa$. We choose $t_{1}=t_{2}=5\,\mathrm{nsec}$. With $\delta
k_{l}=10^{8}\,\mathrm{m}^{-1}$, $n_{l}^{1,2}=10$, for
${^{9}}\mathrm{Be}^{+}$, the gate time is $T=14\,\mathrm{nsec}$;
and for $^{43}\mathrm{Ca}^{+}$, the gate time is
$T=26\,\mathrm{nsec}$, much shorter than the decoherence time of
the qubits.

\textit{Decoherence of the combined system.} In the interacting
system of the ion, the charge qubit and the cavity, decoherence of
any component affects the dynamics of the others. Decoherence of
the ion trap qubit, and of the spatially separated Schr\"{o}dinger
cat states of ion motion, as they appear as part of our gate
dynamics, is well studied \cite{ion_trap_exp_review_wineland}. The
dominant effect is decoherence of the charge qubit due background
charge fluctuations, radiation decay, quasiparticle tunneling
through the junction \emph{etc.} \cite{charge_qubit_schoen}. Here
we concentrate on decoherence introduced by the cavity, which is
the new element in the scheme, in particular the effect of
excitation of quasiparticles in the superconducting transmission
line generated by stray laser photons.

The dissipation of the cavity is
described by a resistor $R_{r}$ in series to the inductance $L_{r}$. Following
the two fluid model, the complex conductance of the superconductor is given by
$\sigma(\omega)=\sigma_{1}+i\sigma_{2}$
\cite{jackson_em,tinkham_superconductivity} for $\omega$ lower than the
quasiparticle gap $\Delta$, with $\sigma_{1}=n_{n}e^{2}\tau_{n}/m$
proportional to the density of normal electrons $n_{n}$, and
$\sigma_{2}=n_{s}e^{2}/m\omega$ proportional to the density of superconducting
electrons $n_{s}$. With the penetration depth $\lambda$
(typically of the order of microns),
we obtain $R_{r}=\sigma_{1}L/\lambda\sigma_{2}^{2} b_{0}$.
Without radiation, the quasiparticle density $n_{n}\sim n_{0}\mathrm{exp}%
(-2\Delta/k_{B} T)$ gives negligible resistance, and hence negligible
dissipation ($n_0$ is the total electron density). However,
with stray photons from the ion trap, quasiparticles
are excited which leads to a resistance of
$R_{r}=R_{n}(n_{ex}/n_{0})$, where $n_{ex}$ denotes the number of excited
quasiparticles. The normal state resistance is $R_{n}\sim10^{4}\,\Omega$ with
given parameters.

The cavity loss can be modeled as a bosonic bath that couples to the cavity
which then transfers the fluctuations to the qubits. With an imaginary time
path integral approach\cite{grabert_phys_rep}, we derive the noise spectral
density on the charge qubit:
\begin{equation}
\label{noise_spectral_density}J(\omega)=(\frac{C_{m}}{2C_{t}})^{2}\omega
Z_{eff}(\omega)\coth{(\frac{\hbar\omega}{2k_{B}T})}%
\end{equation}
where the effective impedance $Z_{eff}$ is a capacitor $(C_{r}+C_{m})/4$ in
parallel to the series of the inductor $L_{r}$ and the resistor $R_{r}$.
The noise spectrum for the ion can be derived similarly.

With the fluctuation-dissipation-theorem (FDT), the decoherence rate can be
derived from the spectral density. With $\omega\ll1/\sqrt{L_{r}C_{r}}$, we
have $Z_{eff}\approx R_{r}$. The decoherence rates are then
\begin{equation}
\label{decoherence_charge_motion}\gamma_{r}^{q}\approx\frac{R_{r}}{R_{k}}%
\frac{2k_{B}T}{\hbar}(\frac{C_{m}}{2C_{t}})^{2},\quad
\gamma_{r}^{x}\approx\frac{R_{r}}{R_{k}}\frac
{2k_{B}T}{\hbar}(\frac{x_{r}}{4d_{i}})^{2},
\end{equation}
where $R_{k}=\hbar/(2e)^{2}$ is the quantum resistance
and $x_{r}$ is the spatial displacement of the dipole. Consider the laser
power of mW, and assume the absorbed power to be nW for a duration of
$100\,\mathrm{nsec}$. We have
$R_{r}=R_{n}/10^{5}$. With the temperature $T=100\,\mathrm{mK}$,
$\gamma_{r}^{q}=50\,\mathrm{msec}^{-1}$ and $\gamma_{r}^{x}=5\,\mathrm{sec}%
^{-1}$. This shows that the dominate decoherence is not the cavity loss
compared with that of the charge qubit.

\textit{Discussion.} Combining two drastically different systems
naturally introduces technical questions of compatibility, such as
the coexistence of an ion trap with a cavity and connected charge
qubit. Ions can either be trapped with a Paul or a Penning trap,
i.e. employing strong electric or magnetic fields, while a
mesoscopic charge qubit can not survive a magnetic field exceeding
$\sim0.1$ Tesla and a voltage exceeding $\sim1$ millivolt. In the
case of a Paul trap typically radio frequency fields up to
$250\,\mathrm{MHz}$ are applied which, according to
Eq.~(\ref{int_i_t_c}), couples to the charge qubit via the
capacitor $C_{i}$. For example, trapping a single
$^{43}\mathrm{Ca}^{+}$ ($^{9}\mathrm{Be}^{+}$) ion in a trap of
the size of $\sim20\,\mathrm{\mu m}$ ring diameter (or cap
distance) requires $V_{trap}$ about $30$ - $50\,\mathrm{V}$ at
$100$ - $250\,\mathrm{MHz}$ to achieve a trap depth of about $1$ -
$1.5\,\mathrm{eV}$ with corresponding trap frequencies of $18$ -
$20\,\mathrm{MHz}$. Thus capacitive coupling of the trap's drive
frequency to the endcaps must be carefully compensated for by
using tailored eletronic filter circuits. This is only
schematically indicated in Fig.~\ref{figure_1}, in all
experimental setups higher order filtering is routinely used.
Thus, the voltage couples to the charge qubit is now
$C_{i}V_{i}+C_{ib}V_{ib}\rightarrow0$ where only a small residue
voltage due to imperfect circuitry passes to the charge qubit.
With a residue of $0.1\,\mathrm{V}$ which is far off resonance,
the dynamics of the qubit is not affected significantly. We note
that the balance circuit requires refined electronic filtering and
feedback control circuitry. In the case of a Penning trap, by
using a superconducting thin film that sustains high magnetic
field or by using a cavity geometry that separates the qubit from
the trap, the qubit can coexist with the trap.

\textit{Coupling of two ions via a cavity.} Instead of coupling an
ion to a charge qubit via a cavity, we can also couple two ions,
albeit at the expense of a reduced coupling strength. This
provides an alternative to the standard scenarios of scalable
quantum computing with trapped ions, which are based on moving
ions \cite{large_scale_wineland,cz_gate_2000}. With the geometry
according to Fig.~\ref{figure_1}(a), the ions couple to the ends
of the cavity. The ion-ion Hamiltonian can be derived similar to
above as
\begin{equation}
\displaystyle H_{i-i}=H_{s}^{1}+H_{s}^{2}+\frac{e^{2}}{2(C_{r}+2C_{i})}%
\frac{\hat{x}_{1}\hat{x}_{2}}{d_{i}^{2}}\label{i_t_i}%
\end{equation}
where $H_{s}^{1,2}$ are the Hamiltonian for the two ions defined in
Eq.~(\ref{Hs}), and the trap voltage can be included by replacing $e\hat
{x}_{1,2}/d_{i}\rightarrow$ $e\hat{x}_{1,2}/d_{i}+C_{i}V_{i}$. With an laser
induced dipole of $ex_{r}$, the interaction magnitude is $4\ln{(d_{0}/b_{0}%
)}e^{2}x_{r}^{2}/4\pi\epsilon_{0}d_{i}^{2}L$. Compared with the direct (free
space) coupling between two dipoles with a distance $L$, the interaction is
enhanced by a factor $4\ln{(d_{0}/b_{0})}(L/d_{i})^{2}$. Besides, the coupling
has the advantage of being switchable: by inserting a switch, e.g. a tunable
Josephson junction, in the circuit, the interaction can be turned on and off
in picoseconds. The coupling is in principle scalable by fabricating multiple
connected cavities.

To conclude, we have studied a generic model and protocol for
coupling qubits stored in trapped ions and solid state charge
qubits via a coaxial cavity. The present example illustrates
prospects of combining and interfacing quantum optical and solid
state systems, and may open new routes towards scalable quantum
computing.

{\em Note added:} After completion of this work, we found the
preprint quant-ph/0308145, by A. S. Sorensen and {\it et al},
discussing coupling Rydberg atoms resonantly to superconducting
transmission line.

Acknowledgments: Work at the University of Innsbruck is supported by the
Austrian Science Foundation, European Networks and the Institute for Quantum Information.

\end{document}